\documentstyle[prb,aps,preprint]{revtex}
\newcommand{\be}{\begin{eqnarray}}
\newcommand{\ee}{\end{eqnarray}}

\begin{document}
\draft
\title{Electrodynamics of Nd$_{0.7}$Sr$_{0.3}$MnO$_3$ Single Crystal investigated
by Optical Conductivity Analyses}
\author{H. J. Lee, J. H. Jung, Y. S. Lee, J. S. Ahn, and T. W. Noh}
\address{Department of Physics and Condensed Matter Research Institute, \\
Seoul National University, Seoul 151-742, Korea}
\author{K. H. Kim}
\address{Department of Physics and Astronomy, Rutgers University, Piscataway, NJ 08855}
\author{S-W. Cheong}
\address{Department of Physics and Astronomy, Rutgers University, Piscataway, NJ
08855 \\
and Bell laboratories, Lucent Technologies, Murray Hill, NJ 07974}
\date{\today }
\maketitle

\begin{abstract}
We investigated temperature dependent optical conductivity spectra of Nd$%
_{0.7}$Sr$_{0.3}$MnO$_3$ {\it single crystal}. We found that polishing and
surface scattering effects on the Nd$_{0.7}$Sr$_{0.3}$MnO$_3$ crystal
surfaces could significantly distort the optical responses, especially in
the mid-infrared region. However, oxygen annealing and gold normalization
processes seemed to remedy the distortions. We found that the spectral
weight of Nd$_{0.7}$Sr$_{0.3}$MnO$_3$ in the metallic state might be
composed of a Drude carrier term and a strong incoherent mid-infrared
absorptions. The temperature dependence of the spectral weight suggests that
the electron-lattice coupling should be important in optical properties of
colossal magnetoresistance manganites.
\end{abstract}



\newpage

\section{Introduction}

It has been well known that series of doped manganites with the chemical
formula {\it R}$_{1-x}${\it A}$_x$MnO$_3$ ({\it R }= La or Nd, and {\it A} =
Ca, Sr, or Ba) near {\it x}~$\sim $ 0.3 undergo paramagnetic insulator to
ferromagnetic metal transitions and show colossal magnetoresistance (CMR)
phenomena near the transitions. The coexistence of metallicity and
ferromagnetism in manganites has been explained by the double exchange (DE)
interaction with a strong Hund coupling between {\it e}$_g$ carrier and {\it %
t}$_{2g}$ spins. However, it has been argued that some additional degrees of
freedom, such as lattice,\cite{millis} orbital,\cite{Ishihara} and disorder%
\cite{sheng}, should be included to explain the CMR quantitatively.

Optical measurements have been very useful to investigate the basic
mechanisms of some insulator-metal transitions and associated electronic
structure changes.\cite{okimoto,takenaka,Kaplan,Kim1} There have been
numerous optical investigations on manganites, however, the experimental
data and their interpretations are varied. Especially, there are several
different views on the electrodynamic responses in the metallic states.

One of such difficulties comes from sample preparation methods. For
examples, Okimoto {\it et al}.\cite{okimoto} investigated optical responses
of polished (La,Sr)MnO$_3$ single crystals and reported a small Drude weight
in the metallic state. Later, Takenaka {\it et al}.\cite{takenaka} showed
that cleaved (La,Sr)MnO$_3$ single crystals had optical spectra different
from those of the polished crystals, and claimed that the small Drude weight
in the polished samples should be originated from damaged surfaces during
the polishing process. The problem due to the sample preparation method is
more serious for Nd$_{0.7}$Sr$_{0.3}$MnO$_3$ thin films. Kaplan {\it et al}.%
\cite{Kaplan} reported an optical response of Nd$_{0.7}$Sr$_{0.3}$MnO$_3$
thin film. They could not observe the Drude-like peak even in the metallic
region at low temperature. Later, Quijada {\it et al}.\cite{Quijada}
reported a Drude-like free carrier behavior for an oxygen-annealed Nd$_{0.7}$%
Sr$_{0.3}$MnO$_3$ film. Therefore, careful surface treatments and exact
optical measurements are inevitable to get correct electrodynamical
responses and to understand the basic mechanism of the CMR.

In this paper, we present optical conductivity spectra $\sigma (\omega )$ of
Nd$_{0.7}$Sr$_{0.3}$MnO$_3$ (NSMO) {\it single crystal} for annealed and
polished surfaces. Although {\it dc} conductivity values of the both
surfaces were nearly the same, their $\sigma (\omega )$ were quite
different. Using the oxygen annealing and gold the normalization techniques,
surface strain and surface scattering effects could be removed. Our
experimental data suggested that the low frequency spectral weight in the
metallic states might be composed of a Drude part and a strong incoherent
mid-infrared absorptions. The temperature dependence of the spectral weight
supports that lattice effects are important in NSMO.

\section{Experimental}

The NSMO single crystal was prepared by floating zone methods. Details of
crystal growth and characterizations were described elsewhere.\cite
{Fernandez-Baca} Just before optical measurements, the sample was polished
up to 0.3 $\mu m$ using diamond pastes. After the polishing, we carefully
annealed the sample again in an O$_2$ atmosphere at 1000 $^oC$ for 1 hour.
Figure 1 shows values of temperature $(T)$-dependent dc resistivity $\rho
(T) $ of our NSMO single crystal, which were measured by the conventional
four-probe method. The solid and the dashed lines represent the $\rho (T)$
curves of the polished and the annealed NSMO, respectively. The $\rho (T)$
curves show a typical insulator-metal transition around 198 K. It seems that
overall features of $\rho (T)$ for the polished and the annealed NSMO are
nearly the same except for small deviations around $T_C$. Although such an
annealing procedure does not change $\rho (T)$ significantly, it seems to be
very important for optical properties of the NSMO sample. As demonstrated in
the next section, optical properties of the NSMO crystal became changed
drastically after the oxygen annealing.

Near normal incident reflectivity spectra, $R(\omega )$, were measured in a
wide photon energy region of 5 meV $\sim $ 30 eV and at temperatures between
15 and 300 K. We used a conventional Fourier transform spectrophotometer
between 5 meV and 0.8 eV. Above 0.6 eV, grating spectrometers were used.
Especially above 6.0 eV, we used the synchrotron radiation from the Normal
Incidence Monochromator beam line at Pohang Light Source (PLS). $T$%
-dependent $R(\omega )$ below 6.5 eV were measured using a liquid He-cooled
cryostat. To subtract surface scattering effects from $R(\omega )$, we used
the gold normalization technique\cite{Jung97}: a gold film was evaporated
onto the sample surface just after spectra were taken. Then, the
reflectivity of the gold coated surface was measured and used for
normalizing $R(\omega )$ of the NSMO sample.

Using the Kramers-Kronig (KK) analysis with the normalized $R(\omega )$, we
obtained $\sigma (\omega )$. For this analysis, $R(\omega )$ in the low
frequency region were extrapolated with the Hagen-Rubens relation\cite{Kim97}
for metallic states and with constants for insulating states. To obtain
accurate $\sigma (\omega )$ in the far-infrared region, we also compared $%
R(\omega )$ extrapolated using the Hagen-Rubens relation with those
extrapolated using the Drude model, and we obtained nearly the same results
in the far-infrared region. The $T$-dependent $R(\omega )$ spectra below 6.5
eV were smoothly connected with the reflectivity data obtained at room
temperature from 6.0 to 30 eV, assuming that reflectivity in the high energy
region should be nearly temperature independent. For the frequency region
above our measurements, the reflectivity at 30 eV was extended up to 40 eV
and then $\omega ^{-4}$ dependence was assumed. Using spectroscopic
ellipsometry (SE), we obtained $\sigma (\omega )$ in the region of 1.5 $\sim 
$ 5.0 eV without relying on the KK extrapolation. The $\sigma (\omega )$
obtained by the SE method were quite consistent with those by the KK
analysis. This fact suggests validity of our gold normalization techniques
and the extrapolations used in the KK analysis.

\section{Results and Discussion}

\subsection{Effects of Oxygen Annealing}

Although the oxygen annealing process does not change $\rho (T)$ very much,
it affects the optical responses of NSMO significantly. Fig. 2(a) shows $%
R(\omega )$ for the polished and the annealed surfaces of the NSMO single
crystal. [Note that both of these spectra were corrected by the gold
normalization technique, so differences are mainly due to the oxygen
annealing effects.] In the insulating region above $T_C$, the differences in 
$R(\omega )$ are not significant. However, in the metallic region below $T_C$%
, $R(\omega )$ for the polished surface are much lower in the frequency
region from 2$\times $10$^{-2}$ to 1.5 eV. The dip in $R(\omega )$,
so-called the ''plasma edge'', for the polished surface appears around 0.6
eV, which is much lower than the observed dip around 2.0 eV for the annealed
surface. The differences in the polished and the annealed surfaces can be
clearly seen in $\sigma (\omega )$, which are displayed in Fig. 2(b). After
the annealing, the spectral weight in the infrared region becomes enhanced
considerably. Although Drude-like peaks at the far-infrared region can be
observed at 15 K for both surfaces, $\sigma (\omega )$ of the polished
surface has a smaller spectral weight.

Changes in the optical response due to the oxygen annealing were already
noticed in NSMO thin films. Kaplan {\it et al}. reported that a NSMO film,
whose resistivity peak temperature $T_P$ was located around 180 K, had a
strong mid-infrared peak and no Drude-like free carrier peak.\cite{Kaplan}
Later, the same group reported that $T_P$ of a NSMO film became about 230 K
when it was annealed in an oxygen environment.\cite{Quijada} They showed
that the strength of the mid-infrared peak became weakened considerably and
that the Drude-like free carrier behavior appeared in the oxygen-annealed
NSMO film. We think that the optical response changes in the NSMO film were
related with the dc resistivity changes reported by Xiong {\it et al}.\cite
{Xiong} They investigated the effects of oxygen annealing in NSMO films.
With more annealing, their resistivity values became smaller and $T_P$ moved
to a higher temperature. They suggested that the annealing effects in the
NSMO films could be explained by a change of oxygen content and a variation
of mixed Mn$^{3+}$/Mn$^{4+}$ ratio.

Since $\rho (T)$ of our NSMO single crystal does not change by the annealing
process, the optical response changes observed in Fig. 2 might not come from
oxygen content changes. We think that the differences in optical spectra
should be originated from damages due to the polishing process. Since the
electron-phonon interactions are believed to be very strong in manganites,
the stress applied during the polishing process could damage the surface
region and deform optical responses significantly.

\subsection{Effects of Gold Normalization}

We also found that light scattering from the polished surface was quite
severe in the NSMO sample. Figs. 3(a) and (b) show $R(\omega )$ and $\sigma
(\omega )$ of the oxygen annealed NSMO sample at 15 K, respectively. The
dashed lines are results without any correction, and the solid lines are
results after the gold normalization. $R(\omega )$ in Fig. 3(a) show that
the light scattering effects are not significant below 0.1 eV, but that they
become very severe between mid-infrared and visible regions. The scattering
loss effects become evident in $\sigma (\omega )$. Fig. 3(b) shows that the
spectral weight becomes reduced in most of the frequency region. Especially,
the spectral change below 0.5 eV becomes quite large.

Recently, Takenaka {\it et al.}\cite{takenaka} compared optical reflectivity
spectra of (La,Sr)MnO$_3$ cleavage surfaces with those of polished surfaces,
and they showed that $\sigma (\omega )$ of the cleaved surface have a much
more enhanced Drude weight. They claimed that the damage of the polished
surface should result in a serious distortion in $\sigma (\omega )$. Since
the optical spectra of their cleaved surface [i.e. Fig. 2 of Ref. 5] are
very similar to our results after the oxygen annealing and the gold
normalization processes, it can be suggested that the differences observed
by Takenaka {\it et al.} were likely to be remedied by our experimental
techniques.

\subsection{Spectral Weight in the Insulating State}

In Fig. 4, we show $T$-dependent $R(\omega )$ of the annealed NSMO single
crystal from 5 meV to 10 eV. At 300 K, there are three sharp peaks
originated from optic phonon modes in the far-infrared region. And, broad
peaks above 0.1 eV come from the optical transitions between electronic
levels. With decreasing $T$, the value of reflectivity approaches to 1.0 in
the dc limit (i.e. $\omega \rightarrow $ 0) and the sharp phonon features
become screened by free carrier. Note that there are strong $T$-dependence
in $R(\omega )$ up to 5.0 eV. Similar behaviors were observed in other CMR
manganites.\cite{Quijada,Jung99}

Figure 5 shows $T$-dependent $\sigma (\omega )$ of the annealed NSMO single
crystal, which were obtained by the KK analysis. In the temperature region
above $T_C$, we can see two interesting features: an optical gap near 0.18
eV, and broad peaks around 1.2 and 4.5 eV. As $T$ decreases, a significant
amount of spectral weight between 1.0 and 5.0 eV moves to a low energy
region. Above 5.0 eV, the spectral weight is nearly $T$-independent. Due to
the similarities in peak position and strength to those of charge transfer
transitions in the other manganites,\cite{Jung99,Jung98,Liu} the peak around
4.5 eV can be assigned to a charge-transfer transition between the O 2{\it p}
and the Mn {\it e}$_g$ bands. The large spectral weight change between 2.0
and 5.0 eV can be interpreted in terms of an interband transition between
the Hund's rule split bands, i.e. {\it e}$_g^{\uparrow }$({\it t}$%
_{2g}^{\uparrow }$) $\rightarrow $ {\it e}$_g^{\uparrow }$({\it t}$%
_{2g}^{\downarrow }$) and {\it e}$_g^{\downarrow }$({\it t}$%
_{2g}^{\downarrow }$) $\rightarrow $ {\it e}$_g^{\downarrow }$({\it t}$%
_{2g}^{\uparrow }$), in the DE picture.\cite{Furukawa,Moritomo} [This
notation indicates that the transitions occur between two {\it e}$_g$ bands
with the same spin but different {\it t}$_{2g}$ spin background.] As $T$
decreases below $T_C$, the spins of the {\it t}$_{2g}$ electrons are
pointing along one direction and the interband transition becomes weakened,
which results in the decrease of the spectral weight.

Now, let us focus on the spectral weights in the insulating state. Although
the peak near 1.2 eV appears to be one broad peak, we claim that it should
be interpreted in terms of a two-peak structure: one peak is located near
1.5 eV and the other is located below 1.0 eV. For most manganites with hole
concentrations below 0.4, the 1.5 eV peak could not be seen clearly.
However, there have been several reports which support the existence of the
1.5 eV peak.\cite{footnote1} From doping dependent conductivity studies on
(La,Ca)MnO$_3$, Jung {\it et al.}\cite{Jung98} showed that such a peak
should exist. Machida {\it et al.}\cite{machida} also observed similar peaks
in transmission spectra of (Nd$_{0.25}$Sm$_{0.75}$)$_{0.6}$Sr$_{0.4}$MnO$_3$
and Sm$_{0.6}$Sr$_{0.4}$MnO$_3$ films. And, to explain optical spectra of Nd$%
_{0.7}$Sr$_{0.3}$MnO$_3$, La$_{0.7}$Ca$_{0.3}$MnO$_3$, and La$_{0.7}$Sr$%
_{0.3}$MnO$_3$ thin films, Quijada {\it et al.}\cite{Quijada} assumed the
existence of the 1.5 eV peak.

Moreover, if the broad peak around 1.2 eV is interpreted as one optical
transition related to small polaron absorption, there will be some
discrepancies between the optical data and other experimental data. For
examples, recent transport measurement shows that a small polaron activation
energy should be about 0.15 eV,\cite{Zhou} which is smaller than 0.3 eV.\cite
{footnote2} When we accept the double-peak structure, we can present a
schematic diagram for the spectral weight in the insulating state, which is
shown in Fig. 6(a). As shown in this diagram, there are four main peaks
below 5.0 eV: (i) a small polaron absorption peak below 1.0 eV, (ii) a peak
centered around 1.5 eV, (iii) a broad peak centered around 3.5 eV, which is
originated from interband transitions between the Hund's rule split bands,
and (iv) a peak due to charge transfer transitions between the O 2{\it p}
and the Mn {\it e}$_g$ bands.

\subsection{Spectral Weight in the Metallic State}

Figure 7 shows detailed $T$-dependent $\sigma (\omega )$ below 2.0 eV. The
solid square, triangle, and circle represent the measured dc conductivities
of 15, 120, and 180 K, respectively. Above $T_C$, we can find three optic
phonon modes. With cooling, the optic phonon modes become screened by free
carrier, but their signature can be observed even at 15 K. As $T$ decreases
below $T_C$, the spectral features above 0.8 eV decrease, and those below
~0.8 eV increase significantly. The increasing spectral weight below $T_C$
seems to form an asymmetric band, whose peak position moves down with
decreasing temperature. Even at 15 K, $\sigma (\omega )$ seem to make a
downturn around 0.1 eV, and the corresponding dc conductivity is higher by
about 1000 $\Omega ^{-1}$cm$^{-1}$ than $\sigma (\omega =0.1$ eV$)$. Similar
spectral weight changes have been commonly observed in other manganites.\cite
{Quijada,Boris,Kim2}

It should be noted that reported behaviors of the broad peak structures in
the low frequency regions (especially, below 0.5 eV at $T\ll T_C$) are
varied quite a lot in earlier works. Okimoto {\it et al.}\cite{okimoto}
measured optical responses of (La,Sr)MnO$_3$ single crystals, and claimed
that the corresponding Drude peaks are very weak and sharp. Similar
behaviors were observed in La$_{0.7}$Ca$_{0.3}$MnO$_3$ polycrystalline
samples by Kim {\it et al.}\cite{Kim2} However, Quijada {\it et al.}\cite
{Quijada} investigated $\sigma (\omega )$ of annealed Nd$_{0.7}$Sr$_{0.3}$MnO%
$_3$, La$_{0.7}$Ca$_{0.3}$MnO$_3$, and La$_{0.7}$Sr$_{0.3}$MnO$_3$ films on
LaAlO$_3$ substrates, and observed broad and strong $\sigma (\omega )$
features at the low frequency region. They attributed these features to
Drude responses and coherent conductions. Boris {\it et al}.\cite{Boris}
measured $\sigma (\omega )$ of a La$_{0.67}$Ca$_{0.33}$MnO$_3$ single
crystal. At 78 K, its $\sigma (\omega )$ showed a very broad Drude-like
behavior which was much stronger than those of polycrystalline samples{\it . 
}Takenaka {\it et al.} also measured $\sigma (\omega )$ of cleaved (La,Sr)MnO%
$_3$ single crystals, and observed a Drude-like peak with a large spectral
weight. The low temperature spectral responses observed by Quijada {\it et
al.}\cite{Quijada} and Boris {\it et al}.\cite{Boris} are similar to our $%
\sigma (\omega )$ data at 15 K, displayed in Fig. 7. It demonstrates that
the low frequency responses of the manganites are very sensitive to the
sample preparation conditions.

One important question which we should address at this point is whether we
should interpret the low-frequency broad absorption feature below 0.5 eV at
15 K as a single component Drude peak. There are three important
experimental facts which we should notice. First, our experimental data
showed the downturn of optical conductivity around 0.1 eV at $T\ll T_C$. The
La$_{0.67}$Ca$_{0.33}$MnO$_3$ single crystal data by Boris {\it et al}.
showed the same behavior. And, the thin film data by Quijada {\it et al.}
showed similar trends, although such behaviors were neglected by the authors
since the downturn regions were close to the low-frequency infrared cutoff
of LaAlO$_3$ substrates. Second, the spectral weight changes of all the
reported experiments, including the NSMO single crystal case displayed in
Fig. 7, cannot be described by a decrease of one spectral component at a
fixed frequency and an increase of a single Drude component near the zero
frequency. Rather, they can be reasonably described by a continual evolution
of an asymmetric peak whose position moves down as the sample changes from
the insulating state to the metallic state. A similar trend was observed for
the room temperature data of the cleaved (La,Sr)MnO$_3$ single crystals,
investigated by Takenaka {\it et al. }As Sr doping increased and the sample
became more metallic, a spectral weight increased in the low frequency
region, which could be described by a continual movement of an asymmetric
peak. Third, some optical data cannot provide reasonable values for
electrodynamic quantities, when a single component Drude peak is assumed.
For example, the single crystalline La$_{0.67}$Ca$_{0.33}$MnO$_3$ data at 78
K indicate mean free path {\it l }$=$ 3.9 $\sim $ 5.2 \AA , which is
comparable to the lattice constant\ $a$. This value is rather too small for
metallic conduction.\cite{Ioffe} On the other hand, assuming the two
components response for La$_{0.7}$Ca$_{0.3}$MnO$_3$ polycrystals, Kim {\it %
et al}.\cite{Kim2} reported that {\it l }$\sim $ 25 \AA\ at 15 K and that 
{\it l} becomes about 4.2 \AA\ near the metal-insulator transition
temperature.

Therefore, we think that the low frequency response should be interpreted in
terms of two components, i.e. a coherent narrow Drude peak and an incoherent
mid-infrared absorption band. Figure 6(b) shows a diagram of $\sigma (\omega
)$ for $T$ $\ll $ $T_C$, which takes account of the two-component
contributions. Other features of the diagram are following: the peak due to
the charge transfer transition is nearly $T$-independent. And, as
temperature decreases below $T_C$, the transition between the Hund's rule
split bands becomes weakened and the spectral weight below 1.0 eV increases
significantly.

As displayed in Fig. 7, the incoherent mid-infrared absorption band is quite
asymmetric and the Drude weight starts just below the characteristic phonon
mode frequency. These optical features are quite consistent with coherent
and incoherent absorptions of a large polaron, predicted by Emin.\cite{emin}
From these observations, it can be argued that a crossover from small to
large polaron states accompanies the insulator-metal transition in NSMO.
Such a crossover from the small to the large polaron was also observed by
other experiments, including Raman scattering\cite{yoon} and pulsed PDF
measurements.\cite{louca} In this polaron crossover scenario, the strong
oxygen annealing dependence of the optical responses in the metallic states
might be explained. Just after the polishing, the surface region should be
damaged quite significantly, which blocks the motion of incoherent polaron
motion. However, after the annealing, such stress-induced damages could be
repaired. More detailed investigations are needed to clarify this scenario.

\subsection{Temperature Dependence of Low Energy Spectral Weight}

To get a further understanding on the low energy spectral weight, we
evaluated the effective number of carrier, $N_{eff}(\omega _C)$, below a
cutoff frequency, $\omega _C$:

\begin{equation}
N_{eff}(\omega _C)=\frac{2m_e}{\pi e^2N}\int_0^{\omega _C}\sigma (\omega
^{\prime })d\omega ^{\prime }{\rm {,}}
\end{equation}
where $m_e$ is a free electron mass and $N$ represents the number of Mn
atoms per unit volume. For actual estimation, the value of $\omega _C$ was
chosen to be 0.8 eV, since the low frequency spectral weight below $\sim 0.8$
eV increases as $T$ decreases below $T_C$, as displayed in Fig. 7.

Figure 8 shows the $T$-dependence of $N_{eff}(0.8$ eV$)$. Above $T_C$, $%
N_{eff}$ is nearly independent of $T$. As $T$ decreases, $N_{eff}$ shows an
increasing behavior. In the simple DE picture, the spectral weight of the
Drude carriers should be proportional to $(1+M^2)/2$, where $M$ is the
magnetization. However, measured magnetization data show a very sharp
increase near $T_C$, so a quantitative agreement between $N_{eff}$ and $%
(1+M^2)/2$ could not obtained. Recently, Kim, {\it et al.} investigated
optical properties of (La,Pr)$_{0.7}$Ca$_{0.3}$MnO$_3$,\cite{Kim3} and found
that all of $N_{eff}/T_C$ data for various values of Pr concentration can be
scaled with $\gamma _{DE}(T)$, which represents the temperature dependent DE
bandwidth due to spin ordering within the DE model predicted by Kubo and
Ohata.\cite{Kubo} To explain this scaling behavior, they adopted the
theoretical results\cite{Roder} based on a Hamiltonian including the DE
interaction and Jahn-Teller (JT) electron-phonon coupling terms. They showed
that $N_{eff}\approx t\xi \gamma _{DE}(T)$, where $\xi $ is the polaronic
band narrowing term. The solid line in Fig. 8 represents the predicted
behavior of $\gamma _{DE}(T)/\gamma _{DE}(0)$, which agrees quite well with
the evaluated values of $N_{eff}(0.8$ eV$)$. This agreement suggested that
both the DE interaction and the JT electron-phonon coupling play important
roles in NSMO.

\subsection{Electrodynamic Quantities}

One of the important issues in barely conducting materials is to determine
electrodynamic quantities, such as scattering rate $1/\tau $, carrier
density $n$, and effective mass $m^{*}$. The mean free path $l${\it \ }and
dc conductivity $\sigma $ can be written in terms of the electrodynamic
quantities:

\begin{equation}
l=\tau \hbar (3\pi ^2n)^{1/3}/m^{*}{\rm {,}}
\end{equation}
and

\begin{equation}
\sigma =ne^2\tau /m^{*}.
\end{equation}
Assuming hole carrier conduction (i.e. $n=0.3$), {\it l} at 15 K was
estimated to be about 13 \AA . [When we assumed electron carrier conduction
(i.e. $n=0.7$), {\it l} should be about 7 \AA .] As shown in the inset of
Fig. 7, we fitted the free carrier absorption with the simple Drude model
and found that $1/\tau $ $\sim $ 120 $\pm $ 20 cm$^{-1}$ provided a
reasonable fit to our low frequency data. Using this value of $1/\tau $, we
found that $m^{*}/$ $m_e$ $\sim $ 21 $\pm $ 3.

Values of $m^{*}$ could be estimated using other experimental techniques,
such as specific heat measurements. Conduction carriers provide a $T$-linear
term, whose coefficient $\gamma $ can be written as $\pi ^2k_B^2N(E_F)/3$,
where $k_B$ and $N(E_F)$ are the Boltzman constant and density of states at
the Fermi energy, respectively. In the free electron model, the
corresponding linear coefficient $\gamma _0$ can be written as $(4\pi
^3m_ek_B^2V/3h^2)(3n/\pi )^{1/3}$, where $h$ is the Planck's constant.
Therefore, $\gamma /\gamma _0$ should be the same as $m^{*}/m_e$. From
specific heat measurements of Nd$_{0.67}$Sr$_{0.33}$MnO$_3$, Gordon {\it et
al}.\cite{Gordon} reported the values of $\gamma (T=0)$ as 25 mJ/K$^2$mol.
[Note that $\gamma _0$ was estimated to be around 0.9 mJ/K$^2$mol.] However,
they claimed that the value of $\gamma /\gamma _0$, i.e. 26, was too large
for the conduction carrier contribution, and that the coupling between Nd
and Mn spins might enhance the value of $\gamma $.

Note that our estimated value of $m^{*}/m_e$ for NSMO is larger than those
of other manganites. From specific heat measurements, Hamilton {\it et al}.,%
\cite{Hamilton} found that $\gamma /\gamma _0$ was around 5 for La$_{0.67}$Ba%
$_{0.33}$MnO$_3$ and around 8 for La$_{0.8}$Ca$_{0.2}$MnO$_3$. And, from
optical measurements, Kim {\it et al.},\cite{Kim2} reported $m^{*}/m_e$ $%
\sim $ 13 for La$_{0.7}$Ca$_{0.3}$MnO$_3$. In the DE model, $T_C$ should be
proportional to an effective transfer integral $t_{eff}$ for an electron
hopping between Mn ions. And, in the tight binding calculation, $N(E_F)$ is
inversely proportional to $t_{eff}$. Since $\gamma $ is proportional to $%
N(E_F)$, $\gamma $ should also be inversely proportional to $t_{eff}$. From
the fact that $T_C$ of NSMO is quite smaller than those of La$_{0.7}$Ca$%
_{0.3}$MnO$_3$ and La$_{0.67}$Ba$_{0.33}$MnO$_3$, the large value of $m^{*}$
in NSMO seems to be reasonable. This also supports the validity of our
analysis for the NSMO metallic state in terms of a small Drude weight and a
strong incoherent peak.

\section{Conclusion}

We investigated the temperature dependent optical conductivity spectra of Nd$%
_{0.7}$Sr$_{0.3}$MnO$_3$ single crystal. We found that the polishing process
could damage the surface region of the crystal and deform the optical
spectra quite significantly. Using the oxygen annealing process and the gold
normalization technique, such a damage due to the polishing process seems to
be recovered. Based on optical conductivity spectra, we provided schematic
diagrams for spectral weight distributions in the metallic and the
insulating regions. We found that the optical response in the metallic
region should be interpreted in terms of a Drude part and a strong
incoherent mid-infrared absorption.

\section{Acknowledgements}

We thank to N. J. Hur, Dr. H. C. Lee, and Dr. Y. Chung for useful discussion
and helpful experiments. We acknowledge the financial support by the
Ministry Education through Grant No. BSRI-98-2416, by the Korea Science and
Engineering Foundation through Grant No. 971-0207-024-2, and through RCDAMP
of Pusan National University. Experiments at PLS were supported in part by
MOST and POSCO.

\begin{figure}[tbp]
\caption{$\rho (T)$ of the NSMO single crystal. The solid and the dashed
lines represent the resistivity curves of polished and annealed NSMO,
respectively. }
\label{Fig:1}
\end{figure}

\begin{figure}[tbp]
\caption{(a) $R(\omega )$ and (b) $\sigma (\omega )$ for polished and
annealed NSMO.}
\label{Fig:2}
\end{figure}

\begin{figure}[tbp]
\caption{(a) $R(\omega )$ and (b) $\sigma (\omega )$ for gold corrected and
not corrected NSMO. In (b), the solid square denotes the value of dc
conductivity.}
\label{Fig:3}
\end{figure}

\begin{figure}[tbp]
\caption{$T$-dependent $R(\omega )$ of the annealed NSMO with gold
correction.}
\label{Fig:4}
\end{figure}

\begin{figure}[tbp]
\caption{$T$-dependent $\sigma (\omega )$ of the annealed NSMO with gold
correction.}
\label{Fig:5}
\end{figure}

\begin{figure}[tbp]
\caption{Schematic diagrams for $T$-dependent $\sigma (\omega )$ of the NSMO
for (a) $T>T_C$ and (b) $T\ll $ $T_C$, respectively.}
\label{Fig:6}
\end{figure}

\begin{figure}[tbp]
\caption{Magnified $\sigma (\omega )$ of NSMO up to 2.0 eV. The solid
square, triangle, and circle represent the dc conductivity values at 15,
120, and 180 K, respectively. In the inset, $\sigma (\omega )$ at 15 K are
fitted with the simple Drude model.}
\label{Fig:7}
\end{figure}

\begin{figure}[tbp]
\caption{$T$-dependent $N_{eff}$ (0.8 eV) of the NSMO single crystal. The
solid line refers to the behavior of $\gamma _{DE}(T)/\gamma _{DE}(0)$.}
\label{Fig:8}
\end{figure}


\begin{references}
\bibitem{millis}  A. J. Millis, B. I. Shraiman, and R. Mueller, Phys.\ Rev.
Lett. {\bf 77}, 175 (1996).

\bibitem{Ishihara}  S. Ishihara, M. Yamanaka, and N. Nagaosa, Phys. Rev. B 
{\bf 56}, 686 (1997).

\bibitem{sheng}  L. Sheng, D. Y. Xing, D. N. Sheng, and C. S. Ting, Phys.
Rev. B {\bf 56}, R7053 (1997).

\bibitem{okimoto}  Y. Okimoto, T. Katsufuji, T. Ishikawa, A. Urushibara, T.
Arima, and Y. Tokura, Phys. Rev. Lett. {\bf 75}, 109 (1995); Y. Okimoto, T.
Katsufuji, T. Ishikawa, T. Arima, and Y. Tokura, Phys. Rev. B {\bf 55}, 4206
(1997).

\bibitem{takenaka}  K. Takenaka. K. Iida, Y. Sawaki, S. Sugai, Y. Moritomo,
and A. Nakamura, cond-matt/9810035 (unpublished).

\bibitem{Kaplan}  S. G. Kaplan, M. Quijada, H. D. Drew, D. B. Tanner, G. C.
Xiong, R. Ramesh, C. Kwon, and T. Venkatesan, Phys. Rev. Lett. {\bf 77},
2081 (1996).

\bibitem{Kim1}  K. H. Kim, J. Y. Gu, H. S. Choi, G. W. Park, and T. W. Noh,
Phys. Rev. Lett. {\bf 77}, 1877 (1996).

\bibitem{Quijada}  M. Quijada, J. \v {C}erne, J. R. Simpson, H. D. Drew, K.
H. Ahn, A. J. Millis, R. Shreekala, R. Ramesh, M. Rajeswari, and T.
Venkatesan, Phys. Rev. B {\bf 58}, 16 093 (1998).

\bibitem{Fernandez-Baca}  J. A. Fernandez-Baca, P. Dai, H. Y. Hwang, C.
Kloc, and S-W. Cheong, Phys. Rev. Lett. {\bf 80}, 4012 (1998).

\bibitem{Jung97}  J. H. Jung, K. H. Kim, D. J. Eom, T. W. Noh, E. J. Choi,
Jaejun Yu, Y. S. Kwon, and Y. Chung, Phys. Rev. B {\bf 55}, 15 489 (1997).

\bibitem{Kim97}  K. H. Kim, J. Y. Gu, H. S. Choi, D. J. Eom, J. H. Jung, and
T. W. Noh, Phys. Rev. B {\bf 55}, 4023 (1997).

\bibitem{Xiong}  G. C. Xiong, Q. Li, H. L. Ju, R. L. Greene, and T.
Venkatesan, Appl. Phys. Lett. {\bf 66}, 1689 (1995).

\bibitem{Jung99}  J. H. Jung, K. H. Kim, H. J. Lee, J. S. Ahn, N. J. Hur, T.
W. Noh, M. S. Kim, and J.-G. Park, Phys. Rev. B {\bf 59}, 3793 (1999).

\bibitem{Jung98}  J. H. Jung, K. H. Kim, T. W. Noh, E. J. Choi, and Jaejun
Yu, Phys. Rev. B {\bf 57}, R11 043 (1998).

\bibitem{Liu}  H. L. Liu, S. L.\ Cooper, and S-W. Cheong, Phys. Rev. Lett. 
{\bf 81}, 4684 (1998).

\bibitem{Furukawa}  N. Furukawa, Y. Moritomo, K. Hirota, and Y. Endoh,
cond-mat/9808076 (unpublished).

\bibitem{Moritomo}  Y. Moritomo, A. Machida, K. Matsuda, M. Ichida, and A.
Nakamura, Phys. Rev. B {\bf 56}, 5088 (1997).

\bibitem{footnote1}  There are still some debates on the origin of the 1.5
eV peak. Quijada {\it et al.} [Ref. 8] assigned it to an interatomic
transition between the Mn$^{3+}$ levels splitted due to the Jahn-Teller (JT)
distortion. Machida {\it et al.} [Ref. 19] attributed it to an optical
transition related to the JT clusters. Jung {\it et al}. [Ref. 14] assigned
it to an intraatomic transition between JT split Mn$^{3+}$ levels. More
systematic studies are required to get a better understanding on the origin
of the 1.5 eV peak.

\bibitem{machida}  A. Machida, Y. Moritomo, and A. Nakamura, Phys. Rev. B 
{\bf 58}, 12 540 (1998); A. Machida, Y. Moritomo, and A. Nakamura, Phys.
Rev. B {\bf 58}, R4281 (1998).

\bibitem{Zhou}  Guo-meng Zhao (private communication).

\bibitem{footnote2}  It is known that the energy of a small polaron motion
measured by optical techniques is about 4 time larger than the activation
energy measured by transport measurements. If the 1.2 eV peak is the small
polaron absorption peak, the activation energy determined by the transport
measurements should be 0.3 eV. Refer to P. A. Cox, {\it Transition Metal
Oxides}, (Clarendon Press, Oxford, 1992).

\bibitem{Boris}  A. V. Boris, N. N. Kovaleva, A. V. Bazhenov, P. J. M. van
Bentum, Th. Rasing, S-W. Cheong, A. V. Samoilov, and N.-C. Yeh, Phys. Rev. B 
{\bf 59}, R607 (1999).

\bibitem{Kim2}  K. H. Kim, J. H. Jung, and T. W. Noh, Phys. Rev. Lett. {\bf %
81}, 1517 (1998).

\bibitem{Ioffe}  According to the Ioffe-Regal criteria, $l$ $\sim $ $a$ near
metal-insulator transition. For the metallic state, $l$ $\gg $ $a$. Refer to 
{\it The Metallic and Nonmetallic States of Matter}, edited by P. P. Edwards
and C. N. R. Rao (Taylor \& Francis, London, 1985), Chapt. 2.

\bibitem{emin}  D. Emin, Phys. Rev. B {\bf 48}, 13 691 (1993).

\bibitem{yoon}  S.\ Yoon, H. L. Liu, G. Schollerer, S. L. Cooper, P. D.\
Han, D. A. Payne, S-W.\ Cheong, and Z. Fisk, Phys.\ Rev. B {\bf 58}, 2795
(1998).

\bibitem{louca}  D. Louca, T. Egami, E. L. Brosha, H. R\"{o}der, and A. R.
Bishop, Phys. Rev. B {\bf 56}, R8475 (1997).

\bibitem{Kim3}  K. H. Kim, J. H. Jung, D. J. Eom, T. W. Noh, Jaejun Yu, and
E. J. Choi, Phys. Rev. Lett. {\bf 81}, 4983 (1998).

\bibitem{Kubo}  K. Kubo and N. Ohata, J. Phys. Soc. Jpn. {\bf 33}, 21 (1972).

\bibitem{Roder}  H. R\"{o}der, J. Zhang, and A. R. Bishop, Phys. Rev. Lett. 
{\bf 76}, 1356 (1996).

\bibitem{Gordon}  J. E. Gordon, R. A. Fisher, Y. X. Jia, N. E. Phillips, S.
F. Reklis, D. A. Wright, and A. Zettl, Phys. Rev. B {\bf 59}, 127 (1999).

\bibitem{Hamilton}  J. J. Hamilton, E. L. Keatly, H. L. Ju, A. K.
Raychaudhuri, V. N. Smolyaninova, and R. L. Greene, Phys. Rev. B {\bf 54},
14 926 (1996).
\end{references}
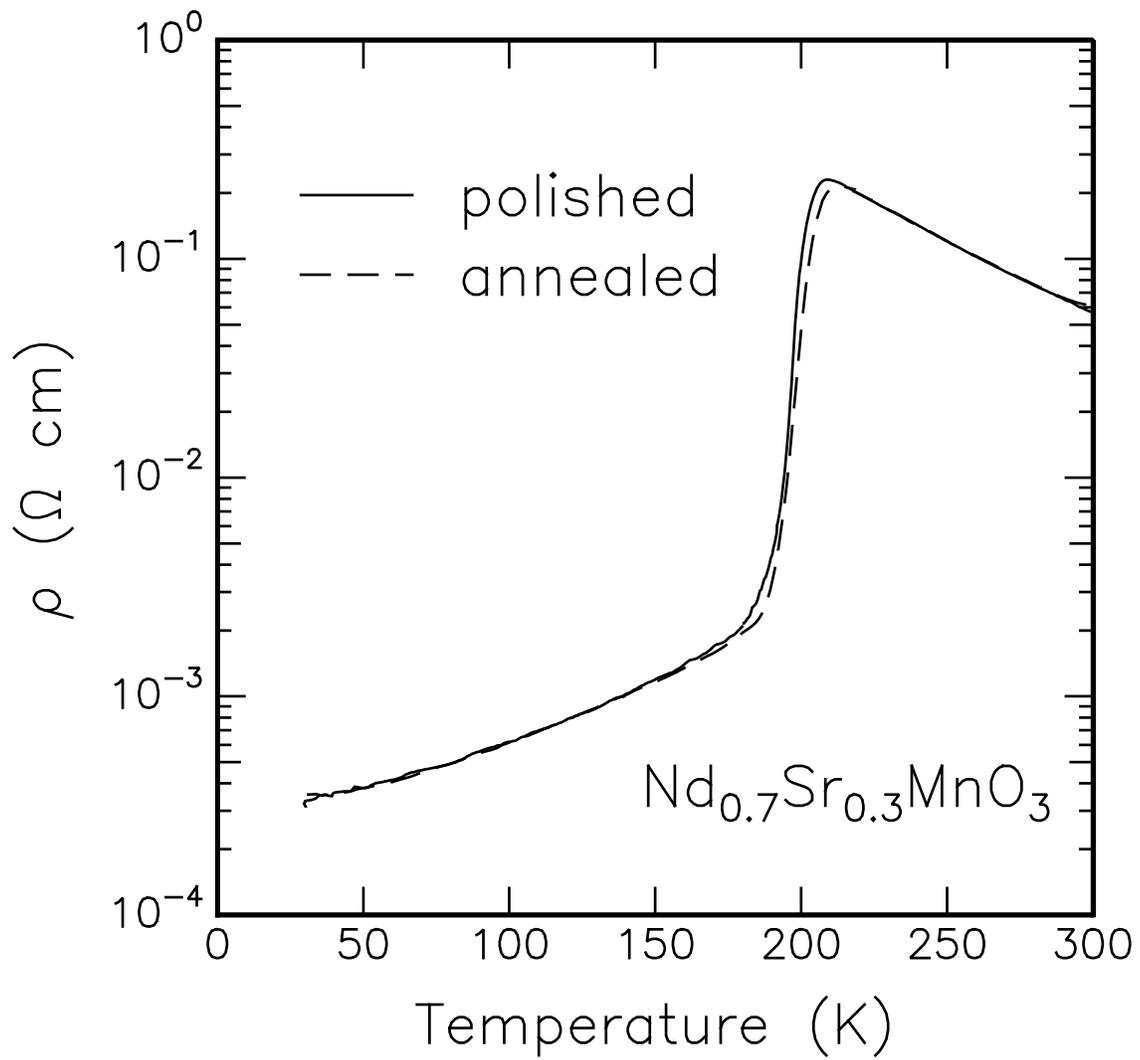
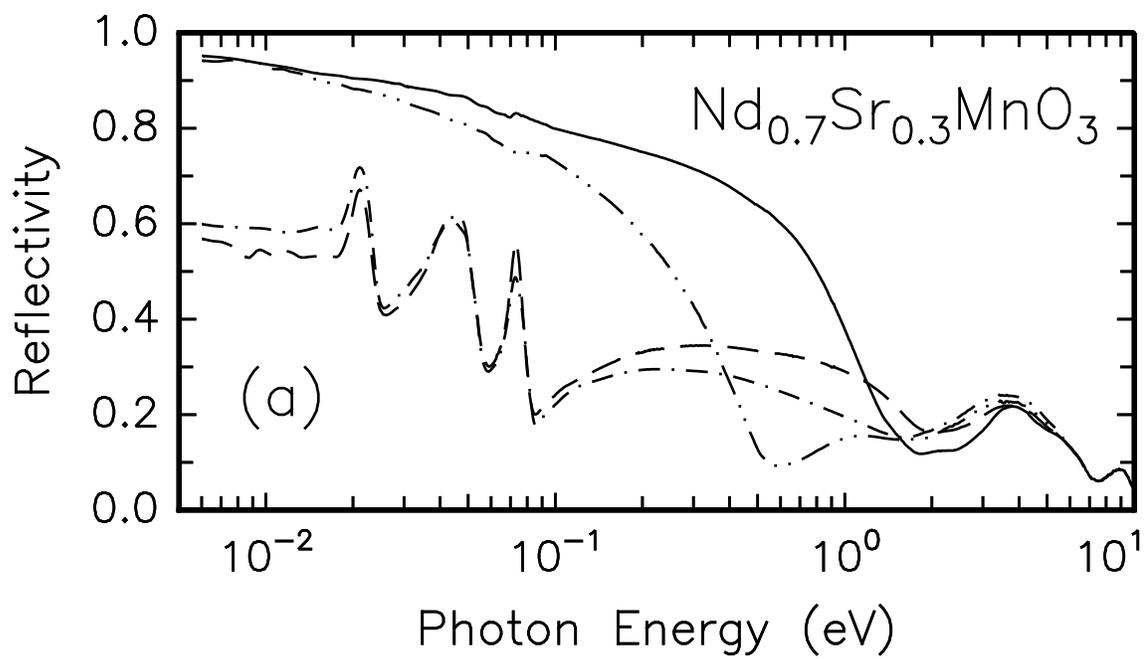
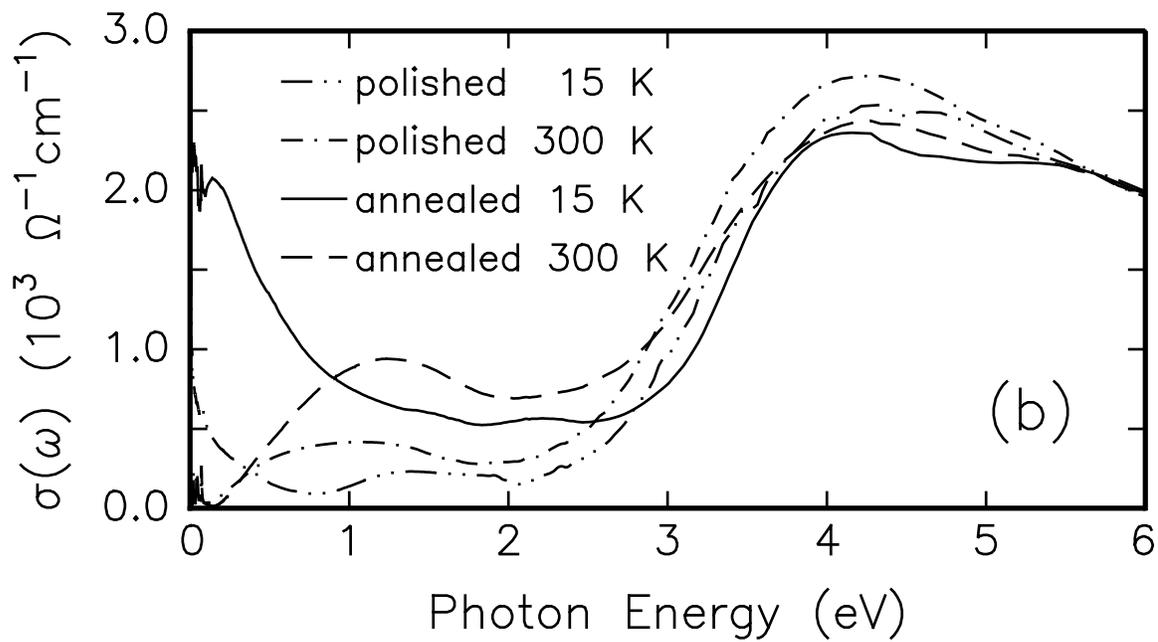
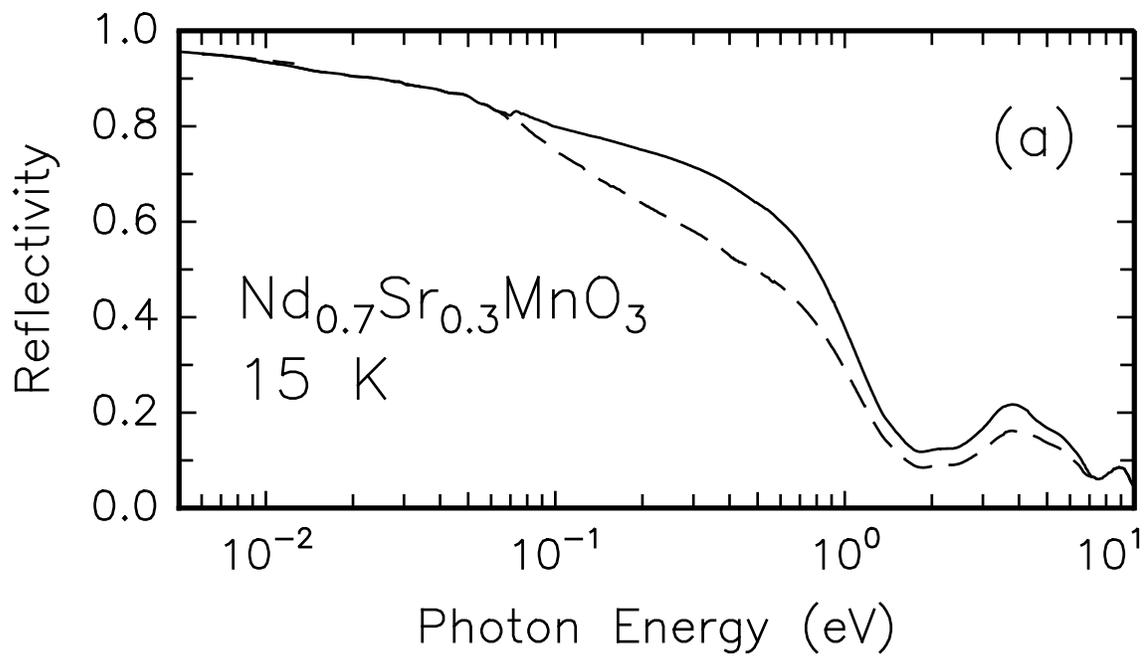
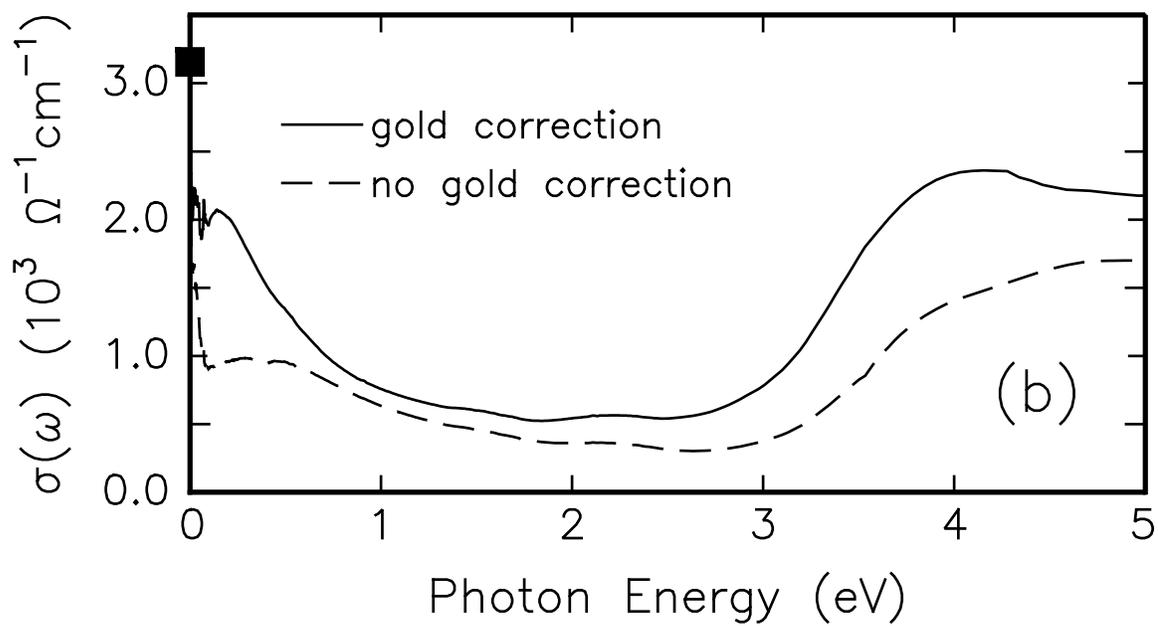
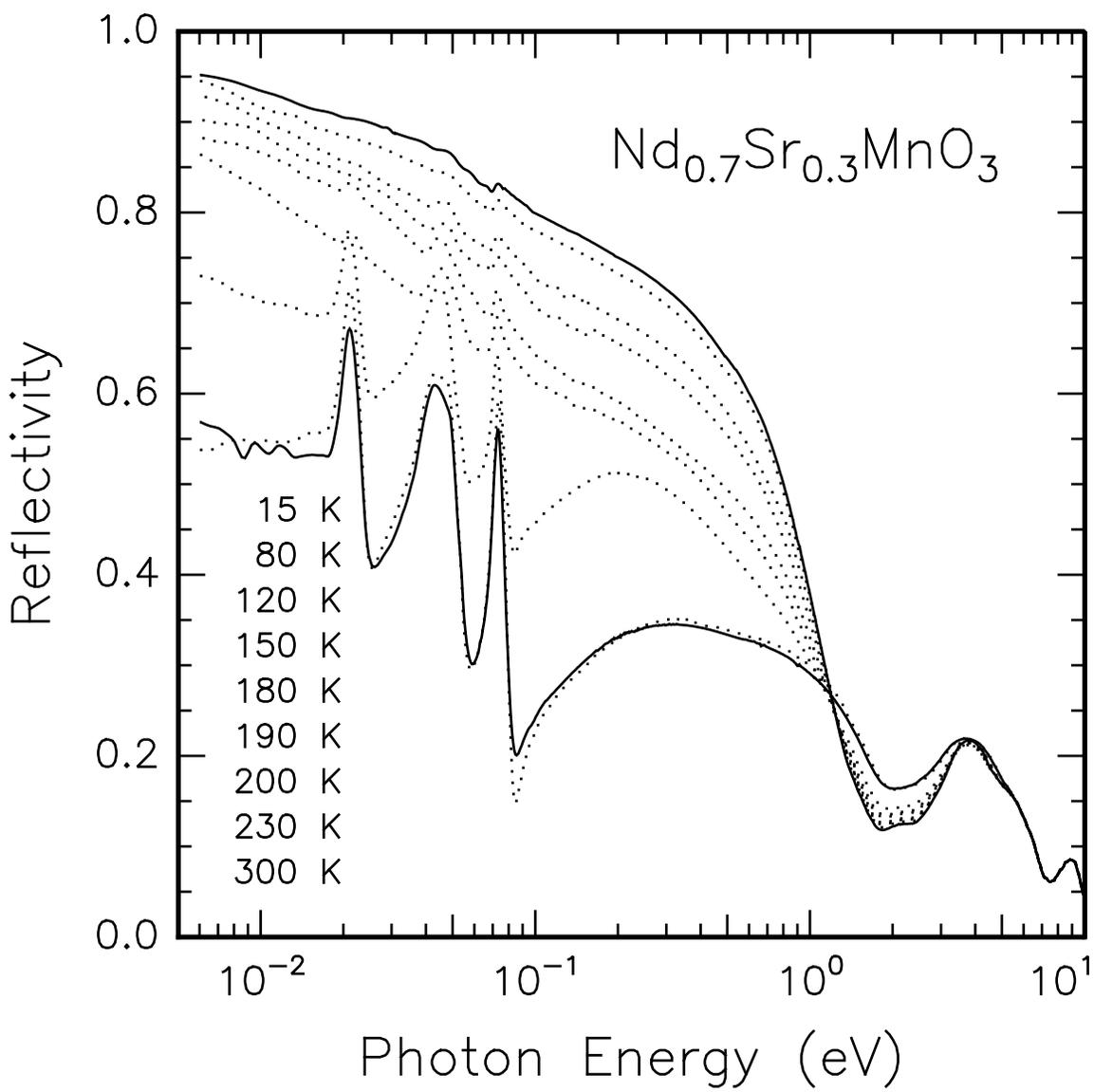
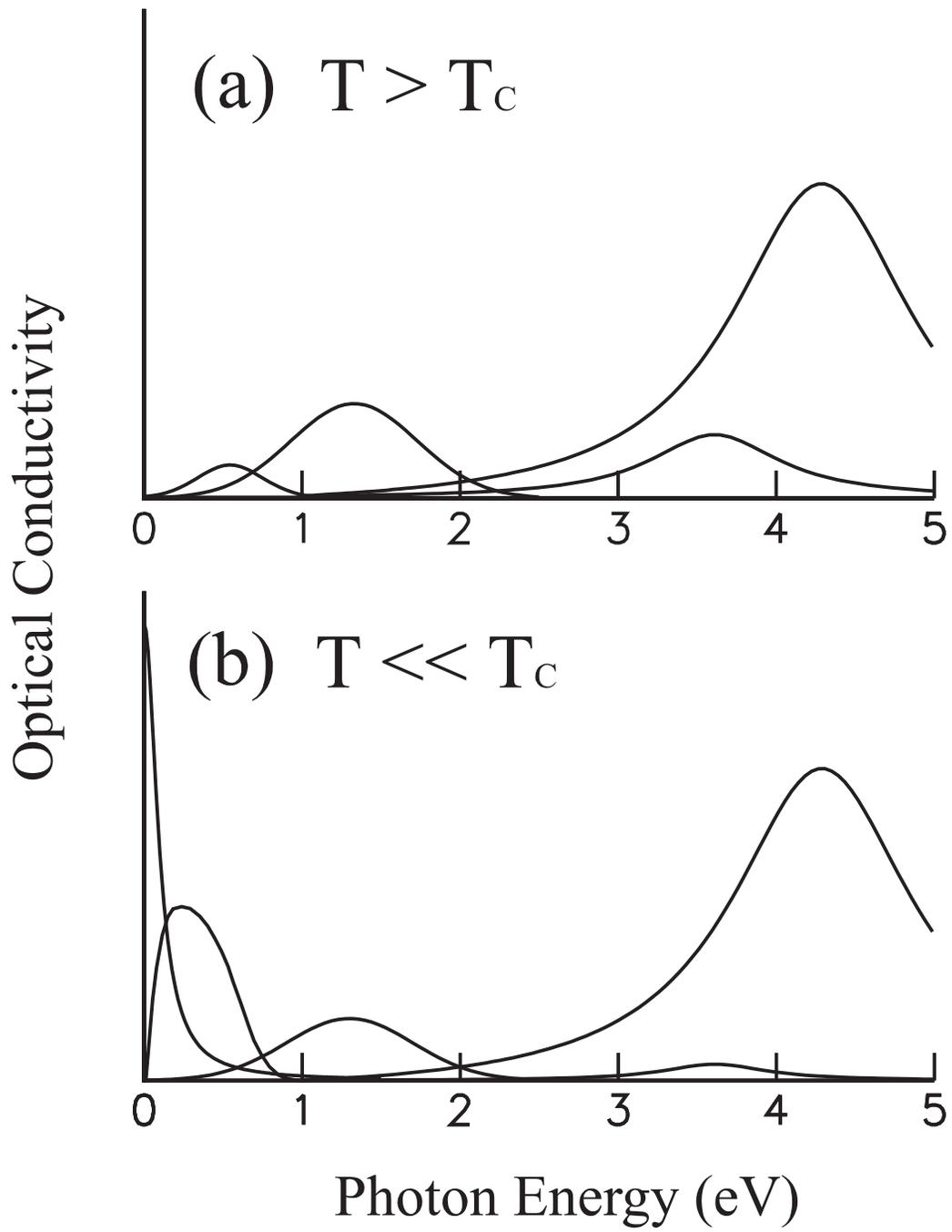
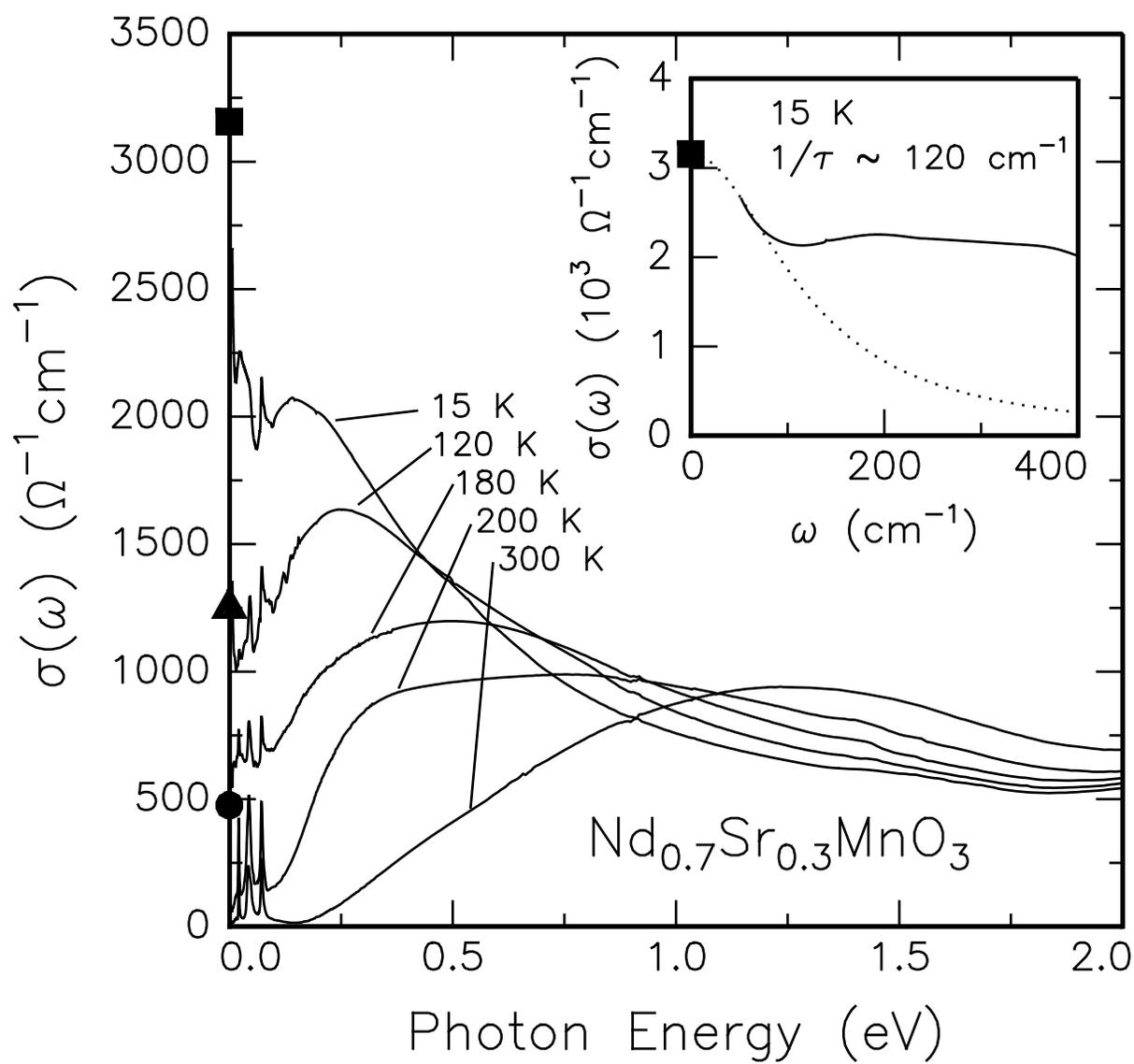
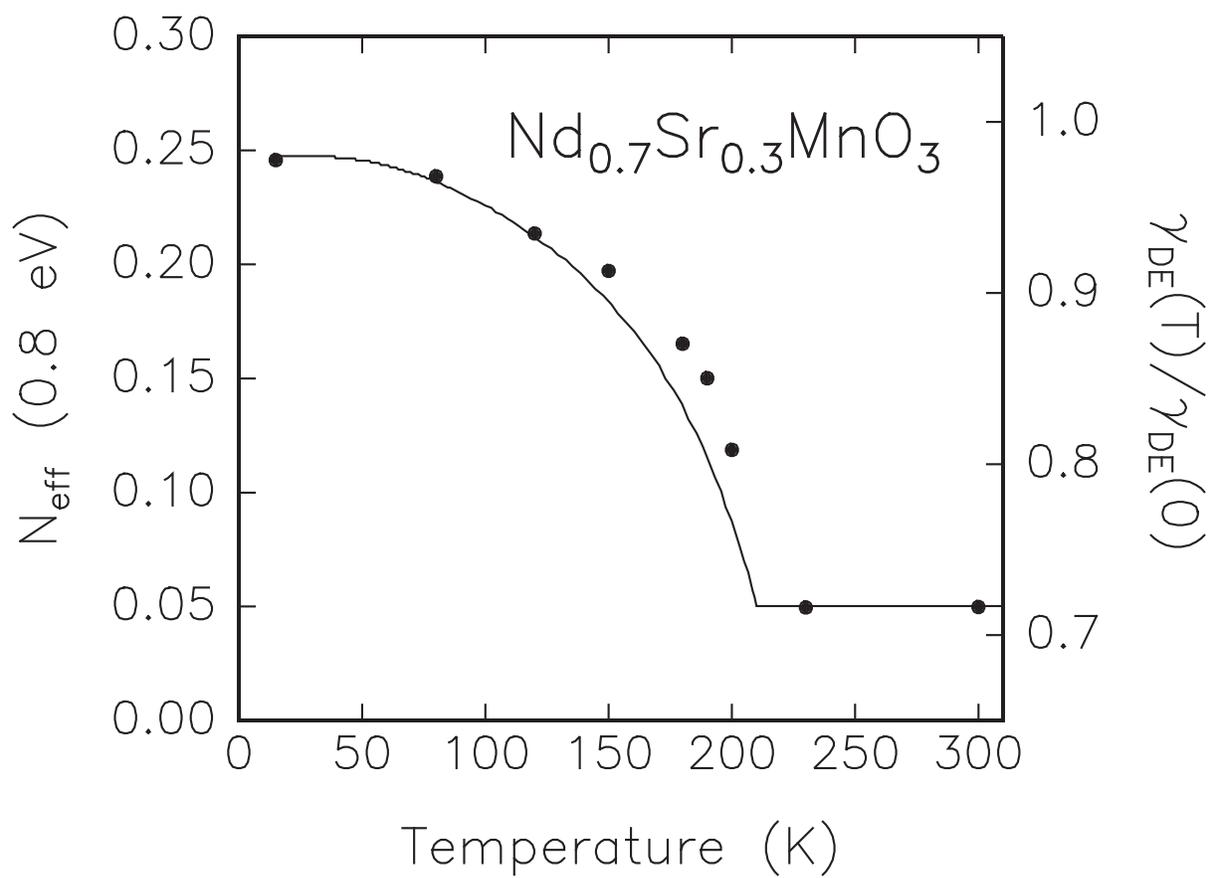
\end{document}